# Next Generation Intelligent Low-Altitude Economy Deployments: The O-RAN Perspective

Aly Sabri Abdalla and Vuk Marojevic
Dept. of Electrical and Computer Engineering, Mississippi State University, USA
Emails: {asa298; vuk.marojevic}@msstate.edu

*Abstract*— **Despite the growing interest in low-altitude economy (LAE) applications, including UAV-based logistics and emergency response, fundamental challenges remain in orchestrating such missions over complex, signal-constrained environments. These include the absence of real-time, resilient, and context-aware orchestration of aerial nodes with limited integration of artificial intelligence (AI) specialized for LAE missions. This paper introduces an open radio access network (O-RAN)-enabled LAE framework that leverages seamless coordination between the disaggregated RAN architecture, open interfaces, and RAN intelligent controllers (RICs) to facilitate closed-loop, AI-optimized, and mission-critical LAE operations. We evaluate the feasibility and performance of the proposed architecture via a semantic-aware rApp that acts as a terrain interpreter, offering semantic guidance to a reinforcement learning–enabled xApp, which performs real-time trajectory planning for LAE swarm nodes. We survey the capabilities of UAV testbeds that can be leveraged for LAE research, and present critical research challenges and standardization needs.**

*Index Terms*—**LAE, O-RAN, UAV, trajectory optimization, vision-based residual network, multi-agent learning, testbeds.**

## I. Introduction

The low-altitude economy (LAE) is an emerging paradigm that leverages different types of unmanned aerial vehicles (UAVs) and artificial intelligence (AI) to transform sectors such as logistics, agriculture, surveillance, and public safety. By enabling low-cost, high-efficiency services, including precision sensing, data collection, and rapid delivery, LAE enhances operational reach in both urban and remote areas. LAE supports real-time feedback and reduced environmental impact compared to traditional transportation and delivery methods; it expands accessibility to critical services, such as emergency response, healthcare delivery, and environmental monitoring. Despite its potential, the large-scale deployment of aerial services faces several challenges, including regulatory restrictions, technological constraints, and infrastructure limitations. A unified framework is required to ensure safe and scalable operations across heterogeneous airspaces and aerial platforms, providing real-time sensing and decision-making, collaborative multi-modal navigation, and adaptive flight scheduling and airspace management [1].

Existing cellular and ad-hoc mesh networks present several limitations for supporting LAE operations, including: (i) lack of mission-conditioned orchestration under partial observability; (ii) single-timescale control loops; (iii) limited network adaptability to new service demands; and (iv) limited coverage and communications scalability in 3D. The open radio access network (O-RAN) introduces a new network architecture that promotes openness, intelligence, virtualization, and interoperability. It disaggregates traditional RAN components and introduces open interfaces, enabling multi-vendor RAN deployment, service scalability, and innovation through AI-driven RAN intelligent controllers (RICs) [2]. The O-RAN near-real time (Near-RT) and Non-RT RICs support closed-loop, multi-timescale optimization and policy-driven orchestration of network resources and user services. This aligns with the requirements of scalable LAE deployments, where real-time decision-making, autonomous coordination, efficient resource utilization, and seamless coexistence of aerial and terrestrial communications are essential.

Enabled by O-RAN, this paper introduces an AI native orchestration framework that fuses semantic priors with online metrics and operates over multi-timescale control loops to stabilize real-time decisions in partially observed 3D airspace. By contrast, existing literature such as [3], [4] operates purely on link-level metrics, lacks mission-conditioned semantics and environment uncertainty awareness, focuses on aerial coverage expansion, and integrates AI into existing systems rather than devising AI native communications and control. Early O-RAN-UAV studies stop at architectural feasibility and key performance indicator (KPI) reporting; they consider single-timescale policy control loops as opposed to Non-RT and Near-RT RIC orchestration and do not account for LAE-specific missions/conditions [5], [6]. To the best of our knowledge, this is the first O-RAN–enabled LAE framework that integrates environmental semantics and mission-conditioned dual-timescale control for LAE swarms.

The rest of the paper is organized as follows: Section II introduces the LAE use cases and the requirements associated with each use case. Section III discusses the potential of O-RAN as a key enabler for LAE operations. Section IV presents a proof-of-concept use case validating the interplay between the RICs for LAE swarm navigation. Section V presents three UAV testbeds enabling LAE research experiments. Section VI discusses critical open issues and research and development directions for expanding O-RAN-enabled LAE capabilities. Section VII highlights the core elements needed for future LAE standardization, and Section VIII offers the concluding remarks.

TABLE I: Representative LAE use cases, performance metrics, and O-RAN-enabled functionalities.

| Application | Theme | QoS Requirements | Key Constraints | Scale | Energy Consumption | O-RAN RIC-enabled LAE Functionalities |
|---|---|---|---|---|---|---|
| Urban Air Mobility | Latency Sensitivity, High Mobility | E2E latency <10 ms, 99.999% reliability, jitter <1 ms, handover failure rate <0.01%, throughput >500 Mbps | 3D trajectory planning, handover, urban canyon effects | Hundreds of UAVs in dense urban zones | High (continuous flight, safety redundancy) | Near-RT RIC/xApp: Real-time beam steering and predictive handover via Signal-to-interference-plus-noise ratio (SINR)/Doppler feedback. Non-RT RIC/rApp: Trajectory clustering and prediction, slice reallocation, swarm-level policy generalization, and long-horizon mobility pattern prediction. |
| Emergency Response & Disaster Recovery | Latency Sensitivity | Latency <20 ms, uplink >100 Mbps, packet loss <0.01%, coverage >95% in disaster zones | Dynamic terrain, intermittent links | Dozens of UAVs in affected region | Moderate to High (hovering + payload operations) | Near-RT RIC/xApp: Adaptive spectrum reallocation and RB scheduling for control and non-payload communication (CNPC) prioritization in degraded conditions. Non-RT RIC/rApp: Federated model training using disaster topologies for policy estimation, autonomous rerouting, and failure-mode reasoning. |
| Smart City Surveillance | Latency Sensitivity, Edge Computing | Video delay <30 ms, encrypted data rate >200 Mbps, uplink reliability >99% | Multi-UAV sync, privacy preservation | Thousands of missions daily city-wide | Moderate (persistent surveillance patterns) | Near-RT RIC/xApp: Uplink stream and inter-UAV communication prioritization and multi-agent control. Non-RT RIC/rApp: Event-driven analytics, adaptive anomaly detection, and slice traffic shaping for QoS enforcement; |
| Transportation and Delivery | High Mobility | Path recalibration <500 ms, delivery accuracy < 1 m, control link availability >99.9%, handoff latency <50 ms | No-fly zones, payload balancing | Tens of thousands deliveries/day | Moderate (frequent launch/ landing cycles) | Near-RT RIC/xApp: Energy-aware network slice allocation and path updates. Non-RT RIC/rApp: Meta-policy learning to optimize energy-latency tradeoffs with policy transfer for energy and congestion-aware routing. |
| Infrastructure Inspection | High Mobility | Continuous video feedback <100 ms, control link latency <20 ms | Continuous data acquisition of large infrastructure, instability in close proximity | Depending on the asset type and mission urgency | Low to Moderate (hover-focused operations) | Near-RT RIC/xApp: Proximity-triggered fast handover/fallback, obstacle-aware beam/null-steering and power control. Non-RT RIC/rApp: Digital twin policy training for trajectory planning, slice pre-provisioning, and anomaly-pattern mining. |
| Precision Agriculture | Edge Computing | Task-to-decision delay <1 s, update rate 10–30 min, rural connectivity >90%, link margin >10 dB | Rural coverage, battery endurance | Field-level deployments | Low (intermittent sensing) | Near-RT RIC/xApp: Wide-area coverage extension via on-demand UAV relaying, burst-mode uplink scheduling for periodic sensing. Non-RT RIC/rApp: Season-aware mission planning and field-level deep learning for crop/soil modeling and analysis. |

## II. Representative LAE Applications and Technical Requirements

LAE spans multiple industry sectors and applications, each with distinct operational goals, quality of services (QoS) requirements, and constraints, which are captured in Table I for representative LAE use cases along with performance metrics, and O-RAN enabled functionalities. The QoS requirements are derived from UAV use cases and applications as defined by 3GPP standardization working groups [7].

### A. Urban Air Mobility

Urban Air Mobility (UAM) is one of the most demanding LAE scenarios, requiring less than 10 ms end-to-end latency and high reliability ($\geq$99.999%) to ensure safe and autonomous navigation across congested, multi-tier airspaces. This requires predictive mobility models to maintain communications continuity with precise localization of UAVs. Coordinated trajectory optimization must also incorporate dynamic edge-compute offloading for onboard path recalculation and traffic management. At scale, UAM scenarios demand orchestration of hundreds of UAVs, each having a high propulsion energy consumption and needing context-aware policy adaptation.

### B. Emergency Response and Disaster Recovery

LAE nodes for disaster response are deployed under infrastructure-deficient conditions for tasks such as search and rescue, damage assessment, and delivery of essential supplies. These missions require aerial nodes to operate under rapidly evolving conditions, such as limited network coverage and unpredictable mobility patterns. Effective UAV-assisted emergency response needs a network that can provide resilient, low-latency communication links, uplink bandwidth exceeding 100 Mbps, and adaptive fallback strategies including redundant UAV relays for handling live video streaming and autonomous decision-making with minimal ground infrastructure support.

### C. Smart City Surveillance

This use case leverages swarms of high-resolution camera-equipped LAE platforms and edge processing units to enable continuous urban monitoring, crowd analysis, and law enforcement support. These missions take place frequently and necessitate low-delay (<30 ms), and encrypted uplink exceeding 200 Mbps for real-time distributed video processing and cross-agent synchronization. Operating within densely populated areas, smart surveillance LAEs often involve inter-UAV coordination to enable distributed formation control, object tracking, and behavior prediction with federated edge analytics applied for face recognition, anomaly detection, and event prioritization, among others. Compliance with data privacy policies needs anonymization protocols, adaptive bitrate control, and robust encryption.

### D. Transportation and Delivery

Autonomous aerial transportation and delivery operate under stringent localization and tracking constraints, with sub-meter delivery precision, congestion-aware routing, continuous tracking support, and path recalibration deadlines below 500 ms. Given the expected high density of delivery UAVs, collision avoidance becomes a major concern. Onboard energy consumption models must adapt to dynamic flight times, payload weights, and mission urgency, calling for power-aware policy optimization for each delivery node.

### E. Infrastructure Inspection and Maintenance

UAVs can perform detailed inspections of critical infrastructure, such as power lines, pipelines, and towers. Sensing and control signaling must remain robust under uncertainty, with error-resilient packet handling mechanisms and latency not exceeding 20 ms for reliable command-and-control (C2). Visual and other sensor data must be processed either locally or at the edge for structural integrity assessment, defect annotation, and maintenance prioritization, requiring onboard compute and mission-aware resource scheduling.

### F. Precision Agriculture

In agricultural and environmental monitoring contexts, aerial nodes conduct multispectral imaging, soil moisture mapping, and pest detection over wide areas with periodic or event-driven sensing tasks. These missions rely on edge inference and localized decision-making to reduce dependence on Cloud connectivity, especially in connectivity-limited areas. Here, latency constraints are relaxed, but rural coverage reliability should be greater than 90%, where energy efficiency is paramount. Solar-assisted or low-duty-cycle energy recharging strategies are typically employed with adaptive flight path planning and data offloading mechanisms.

## III. O-RAN Enabled LAE

O-RAN establishes a disaggregated, modular, and intelligent wireless network. It enables the dynamic orchestration of spectrum, mobility, and computing resources across dense, mobile, and safety-critical LAE environments.

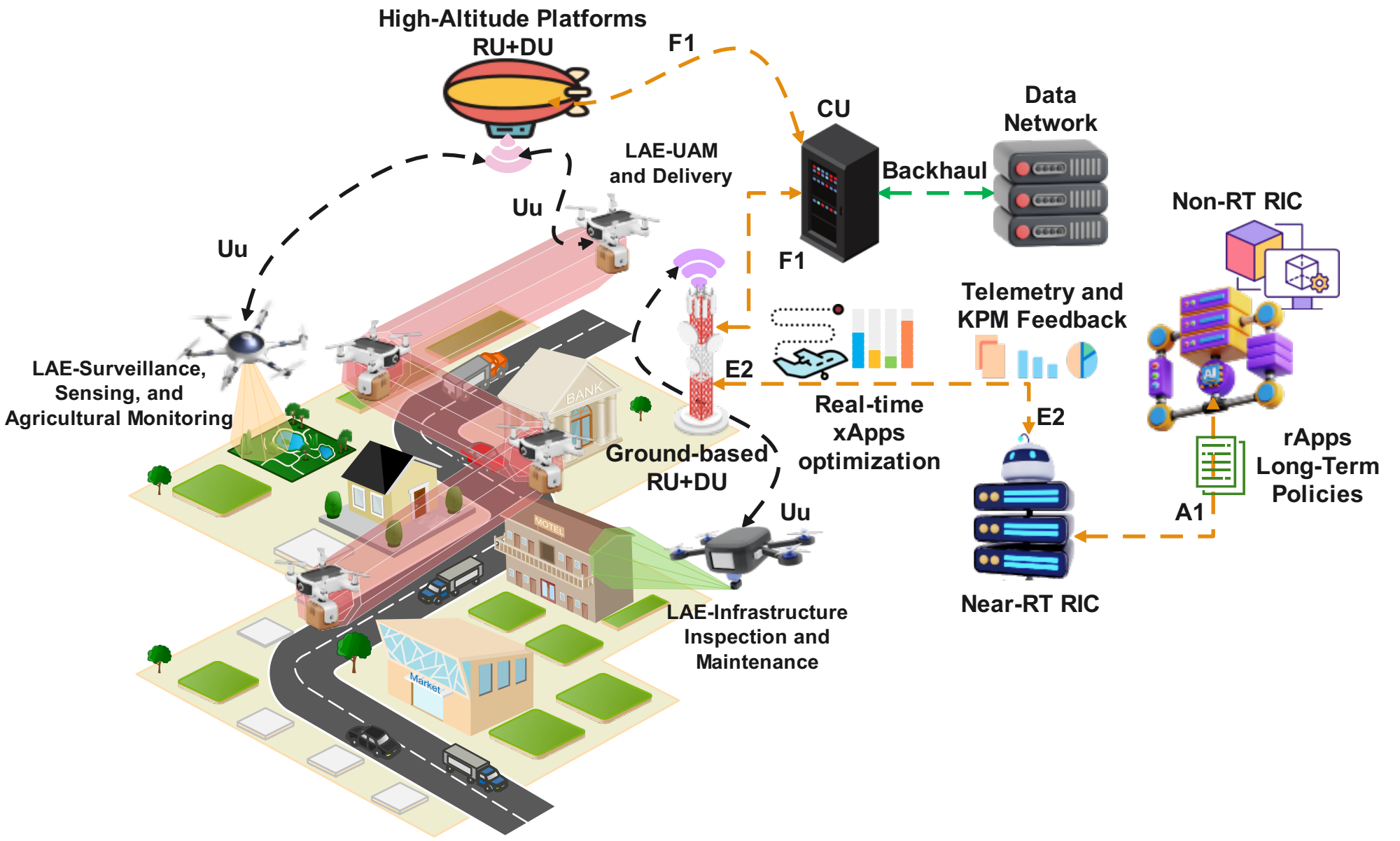

Fig. 1: Terrestrial and non-terrestrial O-RAN-enabled architecture for LAE application support.

### A. Disaggregated RAN Components for LAE

The O-RAN architecture modularizes the traditional monolithic RAN into three logical components, the Radio, Distributed, and Central Units (RU, DU, CU), which are interconnected through standardized open interfaces: the Open Fronthaul (between RU and DU), F1 (between DU and CU), A1 (between Non-RT RIC and Near-RT RIC) and E2 (between Near-RT RIC and DU/CU).

The RU is responsible for radio frequency (RF) signal transmission and reception, beamforming, and analog-to-digital/digital-to-analog conversion. In the LAE context, RUs may be deployed on terrestrial infrastructure and mobile airborne platforms, such as tethered drones serving as aerial RUs that can be strategically placed for improving coverage and capacity in areas of interest.

The DU interfaces with multiple RUs and implements higher Layer 1 and partial Layer 2 processing in such a way to support C2 and time-sensitive payload transmissions. It executes mission-critical functions, including resource block (RB) scheduling, mobility management, and hybrid automatic repeat request (HARQ) feedback handling. Deploying DUs near ground-based edge computing nodes enables tight integration with AI inference.

The CU performs partial Layer 2 and higher protocol operations and manages the connections between the core network and the RAN. The CU aggregates control from multiple DUs and executes centralized resource orchestration and session management. The CU facilitates vertical scaling for LAE deployments by coordinating spectrum and computing resources across different mission profiles, such as high-throughput video surveillance and latency-critical UAM. Policy updates and mobility patterns generated by data-driven algorithms in the Near-RT and Non-RT RICs are implemented at the CU to ensure compliance with network-wide service-level agreements (SLAs) and airspace constraints.

This modular, standards-based functionality separation facilitates network scaling and reconfiguration. It also supports future extensions to include airborne RUs or edge-computing enhanced DUs, among others.

### B. Near-RT and Non-RT RICs for Intelligent Aerial Control

LAE requires dynamic and long-term network optimization for providing evolutionary services in increasingly dense air spaces. The O-RAN RICs form the brain of intelligent control and optimization, enabling advanced decision-making and network programmability. The RICs are logically separated into the Near-RT RIC and the Non-RT RIC, each offering distinct yet complementary control functionalities critical for managing the diverse QoS demands and dynamic behaviors of heterogeneous users, services, and systems.

The Non-RT RIC operates on timescales of seconds and above, and is responsible for policy generation, AI model training, and long-term analytics. Policies and models produced by the Non-RT RIC are distributed via the A1 interface to the Near-RT RIC for near-RT RAN control, enabling continuous adaptation based on long-term trends and strategic objectives. For LAE, the Non-RT RIC aggregates measurements from UAVs and the RAN to identify patterns in data traffic, airspace utilization, interference hotspots, and mission demands. It executes functionalities such as:

- **Policy learning and orchestration:** Historical flight path data and radio maps are used to generate predictive policies for handover, beam switching, and mobility.
- **Model training:** AI models for trajectory prediction, adaptive RB allocation, and energy-efficient routing are trained using federated or centralized learning frameworks.
- **Cross-slice optimization:** Balances network-wide performance across multiple RAN slices, such as low-latency CNPC, high-throughput communications, and high-resolution RF sensing, via intelligent policy adjustment.

Positioned closer to the network edge, the Near-RT RIC operates on a 10 ms to 1 s timescale and directly interfaces with RAN nodes over the E2 interface. It executes time-sensitive, state-driven decisions via xApps, which are lightweight software agents or microservices that can implement AI/ML-based control functions provided by third parties. These xApps continuously ingest LAE measurements and network KPIs, such as throughput and packet error rate, to generate control signals that are relayed to the RAN via E2 Application Protocol messages and from there to the LAE nodes for immediate actuation. The Near-RT RIC enables:

- **Real-time mobility management:** Predictive handover and beam tracking for fast-moving UAVs using Doppler-resilient signal measurements and trajectory estimators.
- **Dynamic spectrum coordination:** Interference-aware resource provisioning for aerial users based on channel measurements and feedback, altitude variation, and user density.
- **Traffic steering:** Balancing network data flows based on priorities, QoS, and congestion awareness.

By enabling coordinated near-RT and non-RT control, the RIC architecture empowers dynamic UAV operations and supports intelligent, mission-adaptive service delivery across multiple LAE use cases such as those captured by Table I. O-RAN thus offers the necessary flexibility, multi-time scale control, and low-latency responsiveness for supporting scalable LAE operations.

Fig. 1 illustrates the proposed O-RAN enabled architecture that supports both terrestrial and aerial deployments using high-altitude platform stations (HAPS). Both deployment options can be implemented with RUs/DUs/CUs, Near-RT and Non-RT RICs, and standardized E2, F1, and A1 interfaces orchestrating dynamic control across diverse LAE verticals.

## IV. Use Case: O-RAN Enabled LAE Vision Aided Path Planning With SINR Awareness

We simulate multi-UAV medical and supply delivery missions across a complex, signal-constrained, and densely obstructed terrain. Each UAV must plan a collision-free path that opportunistically leverages high-SINR positions within the navigation environment to support control signaling and trajectory updates, while ensuring reliable mission execution [8]. This SINR-guided trajectory optimization problem is formulated as a constrained Markov decision process, where each UAV agent selects a trajectory action vector to maximize a long-term cumulative reward subject to collision-avoidance, QoS (SINR), and mission/area constraints. With semantic feature inputs from the Non-RT RIC and real-time SINR feedback obtained from the LAE nodes, the centralized critic and decentralized actors of the multi-agent deep deterministic policy gradient (MADDPG)-based xApp jointly optimize trajectory actions, enabling collision-free and SINR-aware flights. This case study illustrates the interplay between the disaggregated RAN and its AI-native control loops, depicted in Fig. 2.

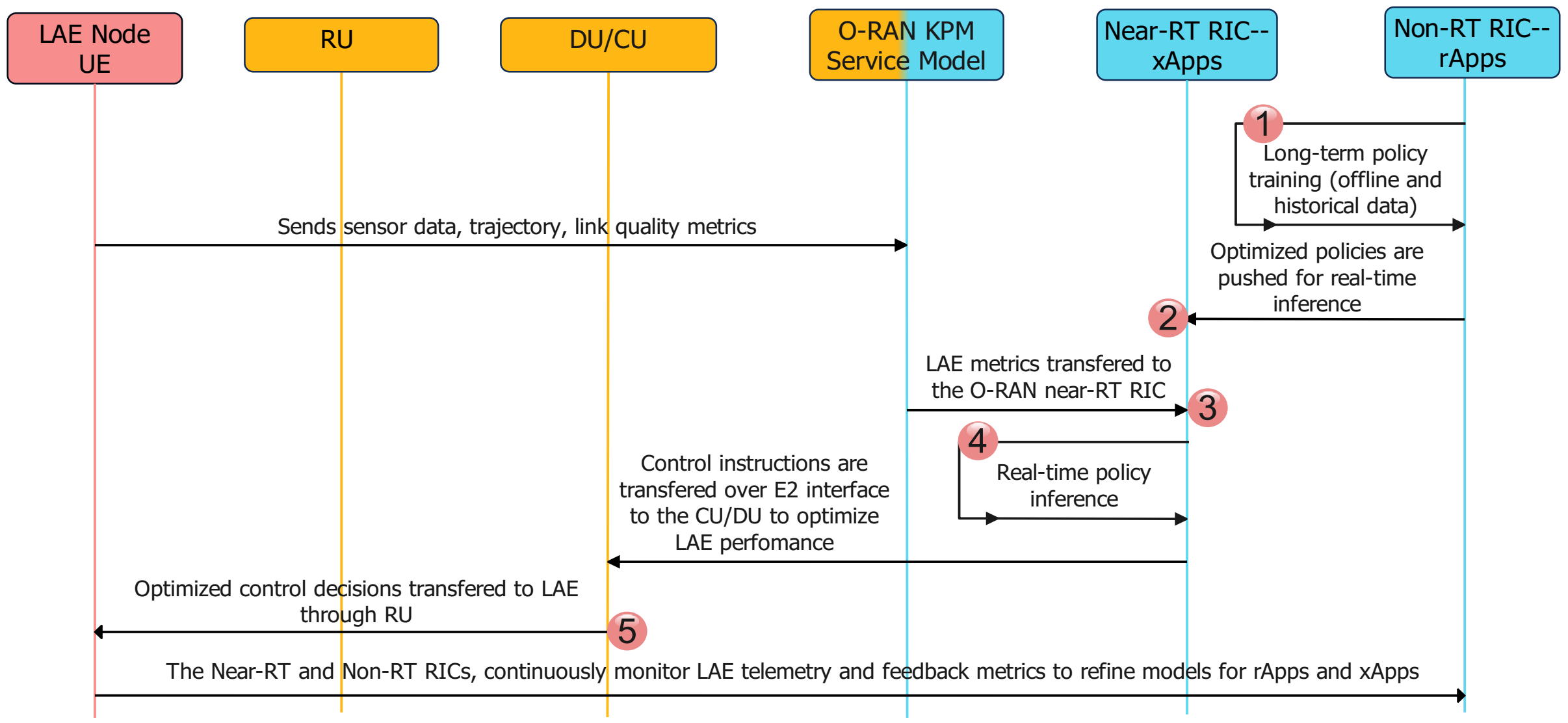

Fig. 2: O-RAN-enabled LAE procedures: flow of actions and O-RAN component interactions.

The operational flow begins with a ResNet-based semantic feature extraction rApp, which is pre-trained offline using high-resolution off-nadir imagery of urban environments as illustrated in Fig. 3(a)-Step (1). Leveraging datasets of real-world aerial building scenes [9], this rApp performs pixel-level segmentation to identify terrain-aware semantic features such as building density, urban occlusion structures, and vertical obstacles. By integrating low-level image edge details with hierarchical context encoding, the ResNet model extracts both shallow spatial textures and deep semantic abstractions [10]. These semantic representations form a spatial knowledge base that is periodically disseminated to the Near-RT RIC via the A1 interface to inform the mission planning and trajectory optimizer xApp (Fig. 3(a)-Step (2)). These semantic features are being gated by a confidence metric, and if the semantic confidence drops due to low resolution, occlusion, or lighting shift, among others, only wireless link metrics will be used for guidance.

The MADDPG xApp performs centralized training and decentralized inference as presented in Fig. 3(b). In each Near-RT interval, the actor for each LAE node continuously collects key performance measurement (KPM) reports, including SINR and UAV position over the E2 interface (Fig. 3(b)-Step (3)) and the confidence-gated semantic features received from the rApp over the A1 interface. It then dynamically optimizes the agent-specific trajectory action vector that adjusts its heading, elevation, and traveling distance to maximize a joint reward function that incorporates QoS, inter-UAV collision margins, distances to target locations, and terrain-aware safety constraints. The reward also penalizes unnecessary altitude changes, unreachable target destinations, and traveling outside the mission area. The centralized critic is trained by minimizing a temporal-difference error with lagged target networks on replayed trajectories collected from the swarm, where each sample includes the agent observation, action, reward, and next observation. The resulting gradients update the decentralized actors via the deterministic policy gradient, after which the updated actors are used for inference (Fig. 3(b)-Step (4)). Optimized control commands are transferred over the E2 interface to the RAN, which forwards them to the LAE nodes (UEs) over the wireless access link (Fig. 3(b)-Step (5)), completing the closed-loop decision cycle of Fig. 2.

Figure 3(c) illustrates the optimized trajectories of the four LAE nodes operating in a densely built urban area. The nodes are initialized at the same starting zone and tasked with reaching spatially diverse target locations to perform delivery missions. The figure shows four UAV trajectory paths, color-coded based on SINR and overlaid on the SINR surface map, derived from a simulated urban wireless propagation model. The environment shown in Fig. 3(c) models building heights and terrain gradients, while the SINR map reflects coverage disparities introduced by occlusion, distance, and fading. The MADDPG converges to navigate the UAV through SINR-enhanced aerial corridors, while maintaining trajectory separation to prevent collisions.

Fig. 4a plots the trajectory of the proposed solution in comparison with three baselines: shortest-path, non-semantic reinforcement learning (RL), and non-SINR semantic RL. The shortest-path flight is the straight line between the start and endpoint. The non-semantic RL policy optimizes the trajectory using SINR feedback but without semantic inputs from the rApp. The non-SINR semantic policy exploits rApp-derived semantics while lacking SINR-awareness. The shortest-path and non-semantic RL trajectories intersect obstacles, thus violating separation constraints, leading to collisions, whereas the proposed path remains collision-free. The non-SINR semantic trajectory avoids obstacles but navigates in low-SINR regions because of the absence of link-quality awareness. Fig. 4(b) plots the SINR of the proposed solution and the three baselines. The proposed solution maintains substantially higher SINR near hot spots and reaches the target with fewer and shorter sub-SINR target (8 dB) intervals compared to the three baselines.

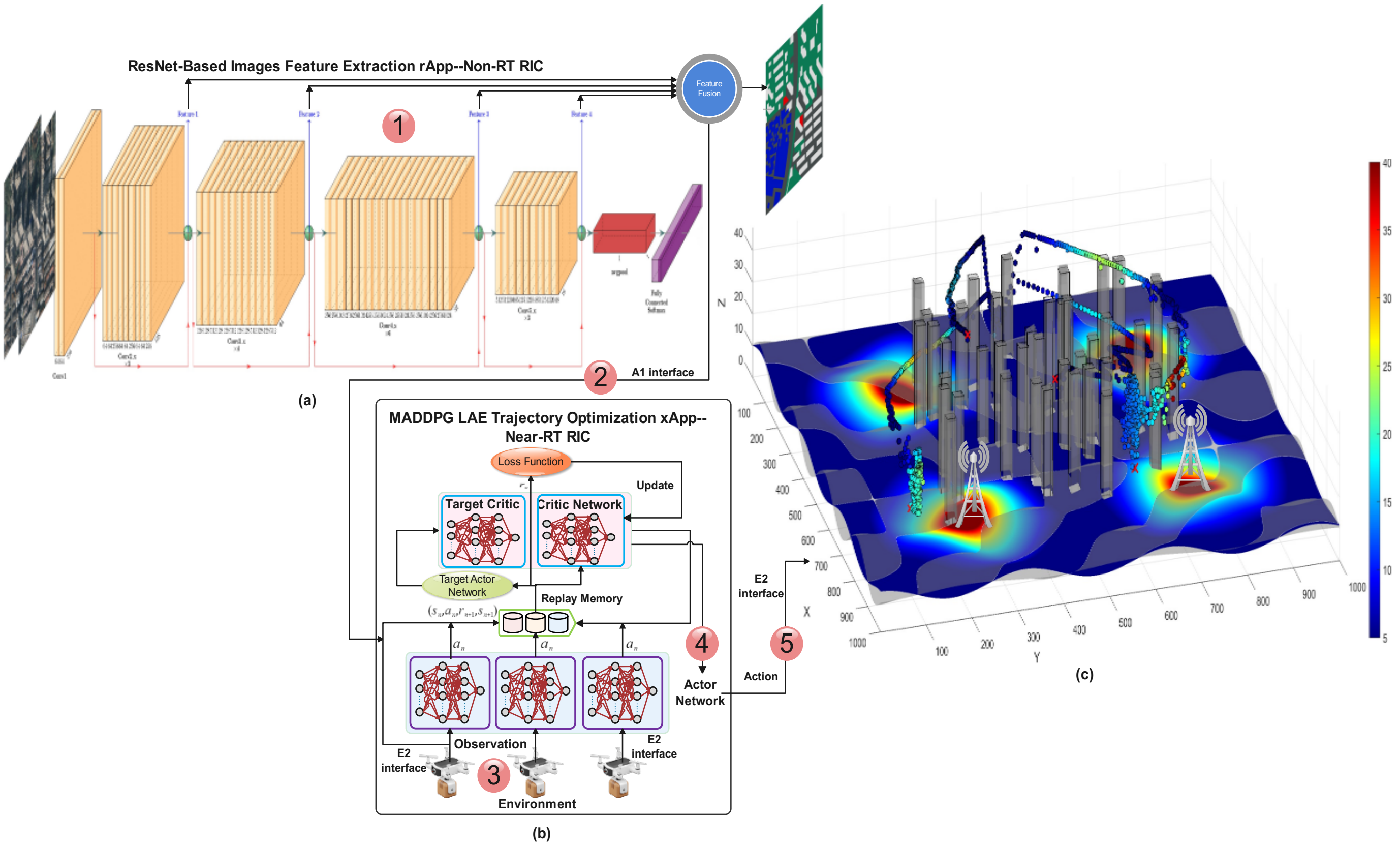

Fig. 3: ResNet-based semantic feature extraction rApp architecture (a), MADDPG trajectory-optimizer xApp (b), and optimized trajectories of four LAE nodes with their SINRs along the way while communicating with the RAN of four cell towers (c).

These results highlight the effectiveness of the proposed O-RAN-enabled LAE framework, which effectively orchestrates semantic reasoning and SINR-aware policy learning in dense, dynamic, and low-altitude UAV deployments. The integration of pre-mission semantic context feature extension and in-mission signaling feedback into a unified decision-making loop enables robust trajectory planning for diverse LAE verticals. While the presented use case targets UAV delivery services, the O-RAN-enabled framework design generalizes to swarm-based surveillance, aerial inspection, and other LEA missions.

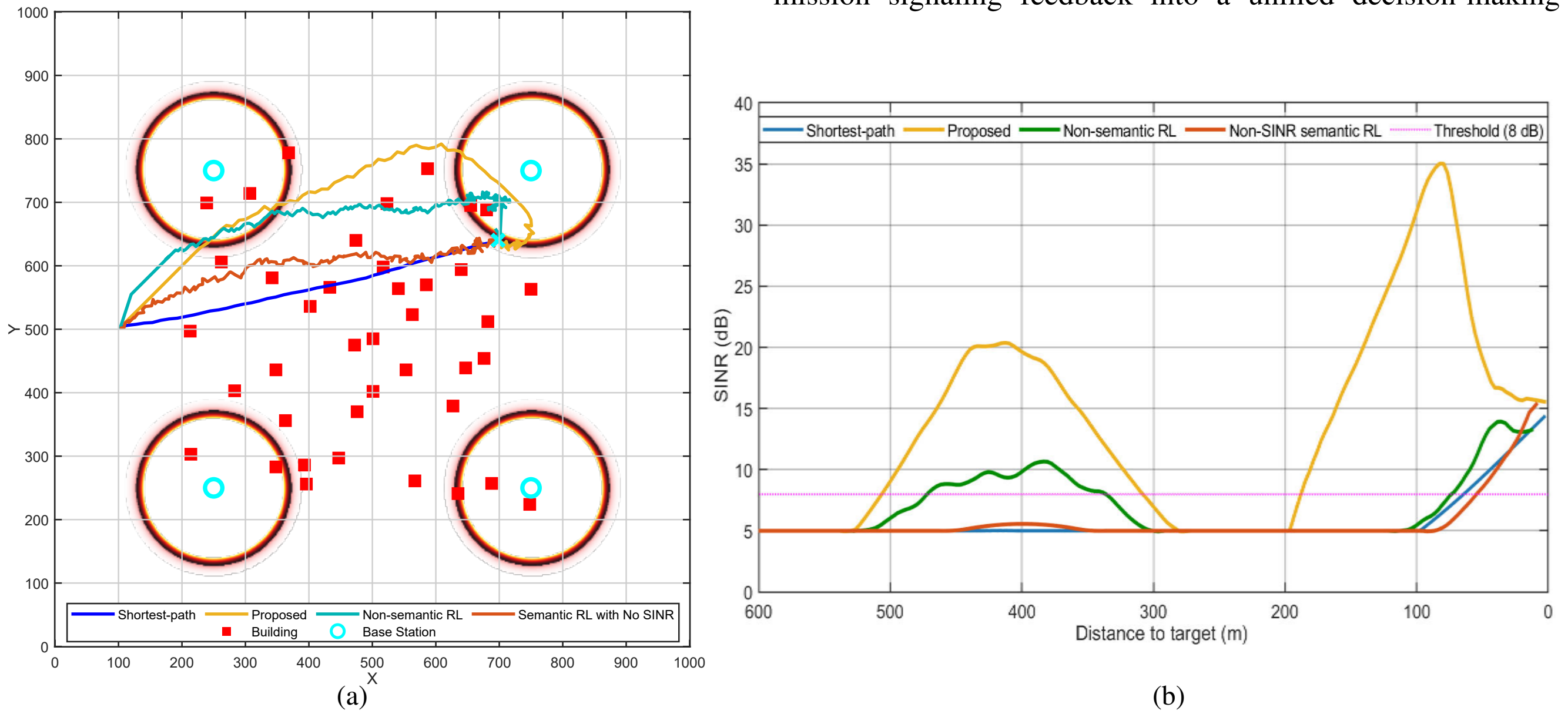

Fig. 4: Proposed trajectory compared to three baseline trajectory solutions for a single UAV (a) and achievable SINR over distance to target for these trajectories (b).

TABLE II: UAV-based testbed capabilities for LAE research and development.

| Capability / Feature | AERPAW | SkyRAN / SkyHAUL | EuroDRONE |
|---|---|---|---|
| Communication Nodes | Aerial UE and gNodeB with OpenAirInterface 5G Standalone SDR deployment | Full 4G RAN/Core on UAVs | Primarily UE / vehicle-to-everything (V2X) role |
| Mobility Awareness | Real-time UAV tracking, handover, edge mobility | Adaptive terrain-aware beam management | Autonomous UAV trajectory with multi-link routing |
| Spectrum Band | Sub-6 GHz, mmWave, C-band, LoRa | 4G Bands + mmWave variants in SkyHAUL | 2.4 GHz / 5 GHz WiFi + Sub-GHz + 4G |
| Digital Twin Integration | RF, I/Q, and protocol-level emulation | Physical UAV-only platform | Physical UAV + software planning, no co-simulation |
| Testbed Scale | Multi-node, Federal Aviation Administration (FAA)-approved flight zone (Lake Wheeler Field Labs) | Multi-UAV mesh + terrain adaptation (Outdoor, terrain-sensitive) | UAV Swarm with heterogeneous radios (Urban air corridor emulation) |
| Potential LAE Use Cases | Urban air mobility, UAV-as-gNodeB, real-time spectrum sharing, edge AI | Disaster recovery, aerial RAN testing adaptive coverage expansion | Cooperative inspection, V2X for UAV corridors, AI planning |
| Accessibility | Available to scientists and industry, remotely accessible, containerized open-source software | Closed industry research and development project, limited public access | Primarily internal research testbed |

## V. UAV Testbeds and Experimental Platforms for LAE Research and Development

There is a pressing need for holistic, system-level testbeds that capture the full complexity of LAE operations. This requires robust experimental platforms that integrate communications, sensing, computing, and control. Three Large-scale UAV communications testbeds that can enable LAE experimentation are described in continuation. Table II summarizes their capabilities.

- **Aerial Experimentation and Research Platform for Advanced Wireless (AERPAW):** AERPAW stands out as an LAE-aligned open research platform. It integrates programmable software-defined radios (SDRs), PX4-based UAVs, and software-programmable base stations and UEs to enable experimentation of aerial communications and control systems [11]. Importantly, it supports both digital twin and real flight experimentation in a controlled airspace with fixed ground nodes, portable UAV-mounted radios, and heterogeneous connectivity (long-range (LoRa), 4G/5G, Wi-Fi). AERPAW has been leveraged to model UAV-to-ground propagation with beam steering and real-world weather impact analysis, handover latency and reliability for mobile base stations, aerial RF source localization and spectrum awareness for dynamic zoning, and O-RAN enabled aerial networking [6]. These capabilities make AERPAW a suitable at-scale experimental platform for validating LAE-native mobility, coordination, and spectrum optimization protocols.
- **SkyRAN/SkyHAUL:** Unlike conventional user-centric UAV testbeds, the SkyRAN and SkyHAUL platforms implement UAV-mounted base station functions, where flying nodes serve as both access points and data aggregators. The SkyRAN system operates with a 4G stack over SDRs, adapting beam patterns and power levels in real-time based on terrain and UE mobility [12]. SkyHAUL creates a self-organizing gigabit wireless backhaul network, integrating multiple UAVs as relay nodes, each capable of dynamic topology adjustments using predictive optimization models [13]. These platforms provide experimental validation for infrastructure-free network coverage in challenging environments.
- **EuroDRONE**: EuroDRONE delivers a compelling framework for multi-radio integration in autonomous LAE settings. It incorporates 2.4/5 GHz Wi-Fi and Sub-6 GHz 4G/5G radios for autonomous UAV swarms to support vehicle-to-infrastructure and vehicle-to-vehicle communications [14]. The platform supports AI-based trajectory planning and heterogeneous delay-tolerant communication strategies for UAM corridors and low-latency delivery services.

## VI. Open Issues and Future Directions

While O-RAN and the proposed framework provide a foundation for supporting LAE operations, several technical gaps must be addressed to support end-to-end autonomy, coordination, and resilience of LAE missions. These challenges span control, data fusion, scalability, and real-time execution.

- **Multi Agent Collaboration Under Partial Observability:** LAE deployments often suffer from obstacles, delayed sensing, or discontinuous connectivity, resulting in partially observable states. Memory-augmented policies, such as long short-term memory, gated recurrent units, and belief-state modeling should be explored for collaborative learning.
- **Scalability and rApp/xApp Resource Contention:** As LAE operations scale to hundreds or thousands of aerial nodes, the O-RAN RICs must concurrently support multiple xApp instances, each processing separate data streams. Current O-RAN implementations lack support for such multi-agent coordination, especially when aerial nodes require overlapping but diverging control policies that can lead to model contention, latency violations, and computing/memory pressure. Scalable orchestration requires development of contention-aware schedulers and admission control mechanisms (with model-cost awareness), policy multiplexing, and distributed xApp clustering and placement with per-agent QoS tags. Meeting latency budgets demands cross-layer co-design of the control plane, KPM report sampling, micro-batching, and compact E2 payloads, coupled with enforced deadline compliance. Future work should pursue computing resources-aware policies with shared backbones, lightweight adapters, quantization and sparsity, dynamic KPM windows, and batch sizes.
- **Digital Twin–Enabled Co-Simulation for LAE:** The lack of emulation frameworks and digital twins that support full-stack co-simulation of flight physics, wireless propagation, and network control in real time present a challenge for LAE research, development, and operations. LAE-focused digital twins that integrate 3D terrain-aware mobility, context-aware propagation, and AI-driven control loops are needed to reduce the costs and risks of flight failures or communications breakdowns. Such digital twins can be integrated within O-RAN's Non-RT RIC for supporting development, integration, and evaluation of learning-based policies in realistic environments using historical and synthetic data before being migrated to the active intelligent controllers in the Non-RT and Near-RT RICs.
- **Low-Latency Semantic Inference:** While the Non-RT and Near-RT RICs do not offer real-time control of network services, many LAE use cases, such as cooperative swarm path planning or task allocation, require control decisions within seconds or less. High-resolution semantic vision and other semantic feature extraction models introduce significant inference overhead. Deploying lightweight models that do not sacrifice spatial detail, semantic accuracy, or reasoning fidelity and that are robust to distributional shifts, which are common in dynamic airspace scenarios, remains an open area of research.

## VII. LAE Standardization

As LAE services evolve from isolated UAV missions to interconnected aerial systems spanning urban logistics, emergency response, and infrastructure inspection, among others, the lack of a unified, scalable standardization framework presents a significant barrier to deployment. Unlike terrestrial 5G systems that benefit from mature standards bodies and vendors, the LAE domain faces fragmentation across airspace regulation, communications, control, and applications.

A core requirement of future LAE standardization is the integration of cross-domain infrastructure spanning aerial, ground, and digital service layers. This cross-domain LAE standardization shall identify key system requirements/functionalities. These functionalities include the physical launch and landing mechanisms, zones, and requirements with maintenance facilities, such as charging, sensor calibration, and vertiport stations and procedures. Moreover, the dynamic aerial operations of LAE need to be standardized within controlled and uncontrolled air spaces via mission-driven authorized flight corridors, altitude levels, and handoff zones. Future LAE standardization also needs to address swarm LAE deployment orchestration with reliable data exchange mechanisms using standard data formats. Most importantly, standardization bodies need to formulate a taxonomy to encode mission intent, risk models, and regulatory constraints into interoperable SLA descriptors that can be enforced across vendors and jurisdictions.

LAE standardization can benefit from the ongoing effort of 3GPP that defines technical specifications for UAV communications over cellular networks, including remote identification and registration of UAVs, cellular-assisted C2 communications, and the coexistence of UAV-based communications with other/terrestrial networks/users [7], [15].

## VIII. Conclusions

This paper advocates for considering O-RAN for supporting the orchestration and optimization of LAE operations. We consider a general LAE use case and introduce a modular, AI-native O-RAN-enabled framework that leverages semantic vision rApps in the Non-RT RIC and multi-agent RL xApps in the Near-RT RIC to support intelligent and adaptive control of UAV swarms in urban and signal-constrained environments. The proposed framework enables mission-critical path planning that jointly considers semantic terrain context, signal strength distributions, and multi-UAV collision avoidance. The use case demonstrates how a ResNet-based rApp can extract building footprints and terrain semantics from imagery, which informs the trajectory optimization xApp. Numerical results highlight the framework's ability to support high-SINR coverage and safe navigation across dense, urban LAE scenarios. This shows the compatibility of the proposed AI-based LAE control architecture with the open, disaggregated, and AI-native O-RAN framework, illustrating how cross-RIC collaboration enables scalable and mission-aware control services. These findings motivate continued research into scalable orchestration, robust AI model deployment, and domain-specific digital twin integration for next-generation O-RAN based autonomous systems. The proposed research and development pathways and the available experimental research platforms open new opportunities for design, deployment, and experimentation of semantic-aware xApps and rApps for diverse LAE applications.

## Acknowledgement

This work was supported in part by the National Science Foundation awards 2120442 and 1939334, and by the Office of Naval Research under Award No. N00014-23-1-2808.

**Aly Sabri Abdalla** (asa298@msstate.edu) is a Research Assistant Professor in the Department of Electrical and Computer Engineering (ECE) at Mississippi State University, Starkville, MS, USA. His research interests are in wireless communication and networking, software radio, spectrum sharing, wireless testbeds and testing, and wireless security with application to mission-critical communications, O-RAN, UAVs, and reconfigurable intelligent surfaces.

**Vuk Marojevic** (vuk.marojevic@msstate.edu) is the Paul B. Jacob professor in the ECE Department at Mississippi State University, Starkville, MS, USA. His research interests include software radios, AI, security, wireless testbeds and testing with application to cellular communications, O-RAN, V2X, mission-critical networks, and unmanned aircraft systems.